\documentclass[aps,prb,twocolumn,superscriptaddress,floatfix,showpacs]{revtex4}
\usepackage{amsfonts}
\usepackage{amssymb}
\usepackage{graphicx}
\usepackage{texdraw}

\usepackage{color}
\newcommand{\GaMnAs}[1]{\mbox{Ga$_{1-x}$Mn$_x$As}}

\begin{document}

\title{Anomalous behavior of  spin wave resonances in \GaMnAs{} thin films}


\author{T. G. Rappoport}
 \affiliation{Department of Physics, University of Notre Dame, Notre Dame, IN
46556}
\author{P. Redli\'{n}ski}
\affiliation{Department of Physics, University of Notre Dame, Notre Dame, IN
46556}
\author{X. Liu}
\affiliation{Department of Physics, University of Notre Dame, Notre Dame, IN 46556}

\author{G. Zar\'{a}nd}
\affiliation{Institute of Physics, Budapest University of Technology and  Economics,
H-1521 Budapest, Hungary}
\author{J. K. Furdyna}
\affiliation{Department of Physics, University of Notre Dame, Notre Dame, IN 46556}
\author{B. Jank\'{o}}\affiliation{Department of Physics, University of Notre Dame, Notre Dame, IN 46556}


\date{\today}

\begin{abstract}
We report ferromagnetic and spin wave resonance absorption
measurements on high quality epitaxially grown \GaMnAs{} thin
films. We find that these films exhibit robust ferromagnetic
long-range order, based on the fact that up to seven resonances are
detected at low temperatures, and the resonance structure survives to
temperatures close to the ferromagnetic transition. On the other hand,
we observe a spin wave dispersion which is {\it linear} in mode
number, in qualitative contrast with the quadratic dispersion expected
for homogeneous samples. We perform a detailed numerical analysis of the
experimental data and provide analytical calculations to demonstrate
that such a linear dispersion is incompatible with uniform magnetic
parameters. Our theoretical analysis of the ferromagnetic resonance
data, combined with the knowledge that strain-induced anisotropy is
definitely present in these films, suggests that a spatially dependent
magnetic anisotropy is the most likely reason behind the anomalous
behavior observed.
\end{abstract}


\maketitle

\section{Introduction}

There has been great deal of interest in elucidating the properties of
${\rm Ga}_{1-x}{\rm Mn}_x{\rm As}$ and other diluted magnetic
semiconductors (DMSs) during the past several years. The intense
research on these materials is partly motivated by the fact that they
hold promise as building blocks of ``spintronic'' semiconductor
devices\cite{ohno98,furdyna00,ohno02}. Indeed, the incorporation of
magnetic properties in
semiconductor heterostructures can, in
principle, lead to the development of new devices that manipulate both
the spin and the charge degrees of freedom of the carriers.  Due to
its compatibility with conventional electronic devices, ${\rm
Ga}_{1-x}{\rm Mn}_x{\rm As}$ is one of the most readily usable alloy
systems for exploring spintronic prototypes. However, these materials
exhibit a series of fascinating strong correlation phenomena that are
not yet fully understood, such as a metal-insulator transition,
field-induced ferromagnetism, and magnetoresistance
\cite{ohno02}. Before any applications become possible, it is
necessary to provide a detailed description of both electronic and
magnetic structure and properties of this class of materials.
Investigating the magnetic properties of the ferromagnetic ground
state in ${\rm Ga}_{1-x}{\rm Mn}_x{\rm As}$, with Curie temperatures
as high as $T_c \sim 160 K$, is therefore especially important.

Given the fact that DMSs are synthesized in film form by using
molecular-beam epitaxy (MBE), ferromagnetic resonance spectroscopy
(FMR) is the most suitable experimental probe for studying the
dynamics of the ferromagnetic order parameter, that also allows for
the spectroscopy of the spin wave excitations. FMR is a powerful
technique to study magnetic properties in magnetic thin
films\cite{farle98} and has already been used by several groups in the
magnetic characterization of the ${\rm Ga}_{1-x}{\rm Mn}_x{\rm As}$
films~\cite{liu03,fedorych02}. The same technique can be used to
obtain the resonance fields of the spin wave modes in a thin film.
Extracting the various magnetic parameters influencing the spin
excitations is essential for gaining complete control of the magnetic
properties of DMS films with an eye on successful future applications,
such as spin injection and manipulation.

The main objective of this paper is to provide a self-consistent
picture of the properties of spin wave excitations based on a
comparison of experimental results and theoretical calculations.
The theory of spin wave resonance (SWR) has been developed four
decades ago by Kittel\cite{kittel58}, who also pointed out that
FMR measurements are capable of detecting several modes of the
magnetic excitations in ferromagnetic thin films. For an external
magnetic field normal to the surface of a homogeneous film he
finds the following resonance condition for spin waves pinned at
its boundaries.

\begin{equation}
H_n=H_{0}-D\left(\frac{\pi}{L}\right)^2 [(n+1)^2-1], \label{eq:kittel}
\end{equation}
where $H_n$ is the external field value (at fixed external radiation
frequency) where the $n$-th SWR mode is observed, $H_0$ is the
magnetic field that corresponds to the ferromagnetic resonance, $L$ is
the sample thickness and the constant $D$ is proportional to the
stiffness constant defined in Sec.~\ref{theory} \cite{spinstiff}.

In this paper we report the observation of such spin wave resonances
in ${\rm Ga}_{1-x}{\rm Mn}_x{\rm As}$. These experiments provide a
direct proof for true {\em long-ranged ferromagnetic order} in
\GaMnAs{}. Surprisingly, the spin waves which we observe exhibit a
somewhat unusual
behavior: We find a spin wave spectrum with $H_n\sim
n$. A recent work by Goennenwein and collaborators\cite{goennenwein03}
reported $H_n\sim n^{2/3}$. Both results are in qualitative
disagreement with Eq.~(\ref{eq:kittel}), which states that for a
homogeneous film the resonance field $H_n$ of the $n^{th}$ mode should
be proportional to $\sim n^2$. We trace back the origin of the
anomalous dispersion to the magnetic properties of ${\rm Ga}_{1-x}{\rm
Mn}_x{\rm As}$ thin films, and find that in order to understand the
anomalous spin wave dispersion, it is {\em crucial} to allow the
magnetic parameters inside the film to depend on the distance from the
interface. In principle, such inhomogeneity in the profile of
magnetization, spin stiffness, or magnetic
anisotropy\cite{goennenwein03} could all result in such an
anomaly. However, our experimental and theoretical results presented
in this paper point towards the presence of uniaxial anisotropy and/or
spin stiffness that depends on the distance $z$ from the surface of
the film. More specifically, we find that our resonance experiments
can be well understood by assuming a uniaxial magnetic anisotropy with
a quadratic dependence on the distance $z$ from the ${\rm
Ga}_{1-x}{\rm Mn}_x{\rm As}/ {\rm GaAs}$ interface . The presence of
strong magnetic anisotropy in ${\rm Ga}_{1-x}{\rm Mn}_x{\rm As}$ and
\mbox{In$_{1-x}$Mn$_x$As} epitaxial films has already been clearly
demonstrated by a series of
experiments\cite{ohno96,abolfath01,dietl01,sasaki02,liu03,fedorych02},
which also indicate that the magnetic anisotropy depends on the
lattice mismatch between the substrate and the DMS layer.

This paper is organized as follows: In Section \ref{experimental} we
present the experimental results for a group of as-grown and annealed
\GaMnAs{} samples and discuss briefly the position, intensity and
linewidth of the observed resonance peaks. In Section \ref{theory} we
present a general spin wave equation of motion that allows for the
variation of the magnetization and magnetic anisotropy along the film
thickness, and discuss our numerical calculation for the position and
intensity of the resonance peaks. Finally, in Section \ref{results} we
compare the experimental results with the theoretical calculations and
discuss the possible explanations for the inhomogeneity of the
magnetic anisotropy.

\section{Experimental Results}
\label{experimental}
 Recently, a systematic study of the fundamental FMR mode has been
 reported for a series of ${\rm Ga}_{1-x}{\rm Mn}_x{\rm As}$ films
 grown on various substrates\cite{liu03}.  The dependence of the FMR
 position on the angle between the applied magnetic field and the
 crystallographic axes of the sample was carefully documented, and
 detailed information has been obtained on the magnetic anisotropy and
 its variation with temperature. The uniaxial and cubic anisotropies
 determined experimentally generally corroborate with earlier
 theoretical predictions~\cite{abolfath01,dietl01}. In this paper the
 same experimental technique is used to study spin wave resonances
 (SWRs) in thin ${\rm Ga}_{1-x}{\rm Mn}_x{\rm As}$ films.

\begin{figure}[b]
\includegraphics[width=6.50cm]{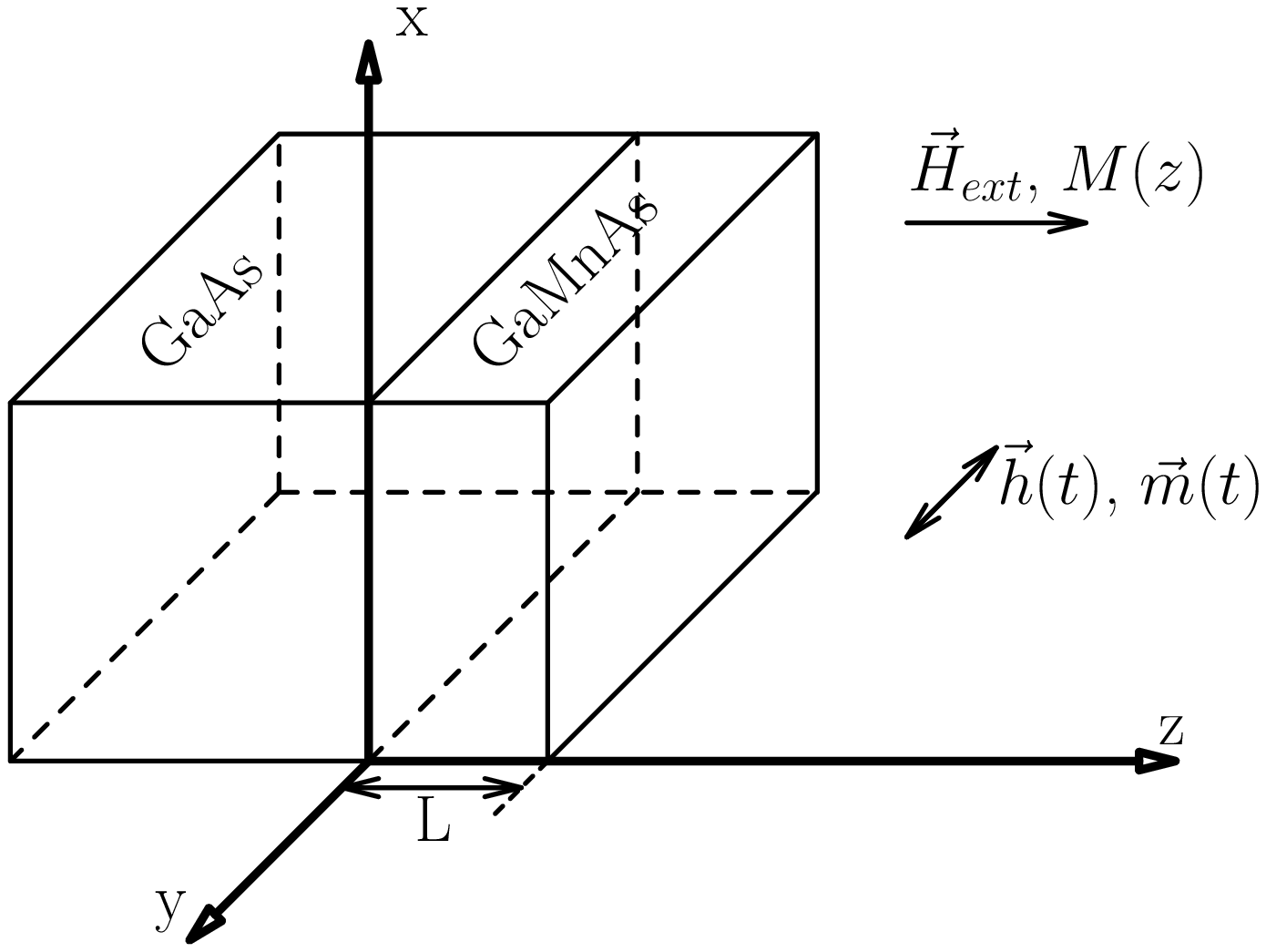}
\caption{Schematic diagram of the experimental geometry. The ${\rm
Ga}_{1-x}{\rm Mn}_x{\rm As}$ film is grown on a thick GaAs
substrate. A large constant magnetic field $\vec{H}_{ext}$ is applied
in the $\hat z$ direction: $\vec{H}_{ext} || \hat z$. This field
produces a large magnetization $\vec{M}(z)$ which points predominantly
in the $\hat z$ direction. Additionally, a small external microwave
magnetic field $\vec{h} \perp \vec{H}_{ext}$ produces a small
harmonically-varying perturbation in magnetization, $\vec{m} \perp
\hat z$} \label{setup}
\end{figure}

Although DMSs have a very low concentration of magnetic atoms, and
these system are often described in terms of
percolation\cite{kaminski02} and impurity band models
\cite{berciu01,fiete03}, it was possible to observe SWR spectra
with up to seven spin wave modes~\cite{sasaki03,goennenwein03},
demonstrating that real long-range order develops in these
materials.

The three  ${\rm Ga}_{1-x}{\rm Mn}_x{\rm As}$ thin films analyzed
in this paper have previously been studied by Sasaki {\it et al.},
who observed both the SWR and the FMR lines. Their initial work
concentrated on the uniform mode (the FMR line), and they used
this feature to investigate the overall magnetic anisotropy in
\GaMnAs{} films~\cite{liu03}. All the samples were grown by low
temperature molecular beam epitaxy (LT-MBE) on semi-insulating
GaAs substrates. The Mn concentration $x=0.076$ was determined by
x-ray diffraction, and the Curie temperature and dependence of
magnetization dependence on applied magnetic field and temperature
were obtained by superconducting quantum interference device
(SQUID) measurements. The FMR and SWR measurements were carried
out using a 9.46 GHz microwave spectrometer.

\begin{table}
\begin{tabular}{||c|c|c|c|c|c|c||}
\hline\hline
  Thickness & Annealed & T & $H_{max}$ & $M$  & $T_c$ & $\Delta H$ \\
   &[nm] & [K] & [Oes] &  $\left[\frac{\rm emu}{{\rm cm}^3}\right]$ & [K] & [Oes]\\
\hline
  & No & 4  & 7820 & 17.9 & 65 & 200 $\pm$ 20 \\
 200  & No & 40 & 7600 & 9.4 & 65 & 170 - 200 $\pm$ 20 \\
  & Yes & 4 & 8700 & 25 & 95 & 120 - 220 $\pm$ 20\\
\hline
  & No & 4 & 8350 & 17.5 & 65 & 230 $\pm$ 20 \\
 150 & No & 40 & 7920 & 9.4 & 65 & 230 $\pm$ 20 \\
  & Yes & 4 & 8820 & 23 & 90 & 180 - 230 $\pm$ 20\\
\hline
  & No & 4 & 8110 & 17.5 & 65 & 150 $\pm$ 20 \\
 100 & No & 40 & 7740 & 9.4 & 65 &  200 $\pm$ 20 \\
  & Yes & 4 & 9050 & 25 & 100 & 110 $\pm$ 20\\
\hline\hline
\end{tabular}
\caption{Experimental parameters extracted from, and used in the
theoretically analysis of, the spectra of three
\mbox{Ga$_{0.924}$Mn$_{0.076}$As} samples of different
thicknesses, pre- and post-annealing. $H_{max}$ is the resonance
field of the highest (ferromagnetic resonance) peak, $M$ is the
static magnetization measured by SQUID, $T_c$ is the critical
temperature, and $\Delta H$ is the linewidth. When the width
varies with the mode number (see text), the range of linewidths is
given.} \label{small_table}
\end{table}

The dc magnetic field $H_{ext}$ was oriented perpendicular to the film
plane (see Fig.~\ref{setup}). As a result of the lattice mismatch
between the substrate and the \GaMnAs{} films, the samples exhibit
compressive strain in the sample plane, which leads to a strong
uniaxial anisotropy. Consequently, in the absence of the external
field the magnetization lies in the plane parallel to the
film. However, the applied static field at which FMR and SWRs are
observed is strong enough to align the magnetization perpendicular to
the sample surface. After measuring the SWR in the as-grown samples,
we annealed them for 60 minutes.  As a result, the Curie temperature
$T_c$ was increased by about $\sim 40\%$ and the magnetization by
$\sim 25\%$. Table~\ref{small_table} summarizes some of the
characteristic properties of three representative ${\rm Ga}_{1-x}{\rm
Mn}_x{\rm As}$ samples. The values listed in the table were extracted
from SQUID, FMR and SWR measurements. Clearly, the annealing process
has a profound effect not only on the critical temperature and
magnetization of the films, but on practically all characteristics of
the the ferromagnetic and spin wave resonances, including the
temperature and thickness dependence of the resonance position and
linewidth. In order to obtain more insight into the behavior of these
materials, we now proceed to evaluate the actual FMR spectra before
and after annealing.

\begin{figure}
\includegraphics[width=8cm]{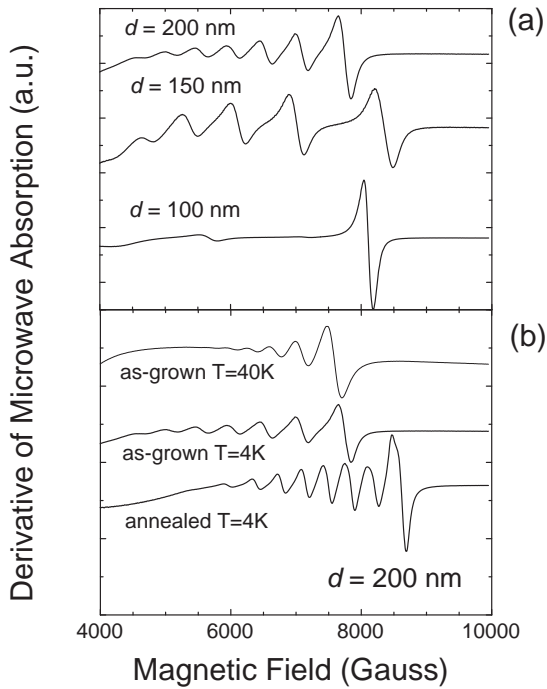}
\caption{\label{fig0}(a) FMR and SWR absorption spectra for as-grown
 \mbox{Ga$_{0.924}$Mn$_{0.076}$As} samples of different thickness
  (100~nm, 150~nm and 200~nm) at \mbox{T=4 K}. (b) FMR and SWR spectra for the 200~nm thick
  \mbox{Ga$_{0.924}$Mn$_{0.076}$As} sample for different
  temperatures, before
  and after annealing.
}
\end{figure}

\begin{figure}
\includegraphics[width=7cm,clip]{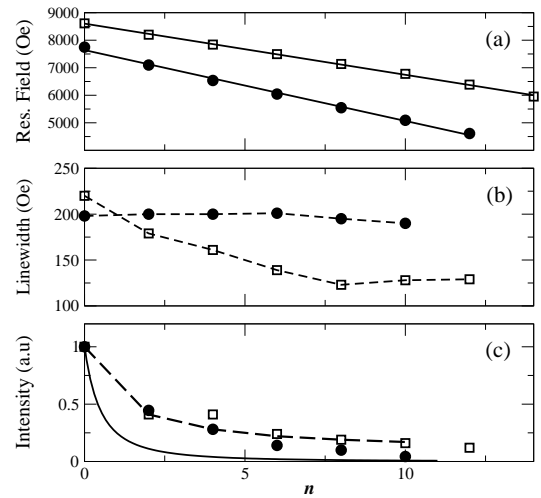} \vskip1truecm \caption{
\label{respositions} Experimentally observed $T={\rm 4K}$ SWR mode
positions (a), linewidths (b) and peak intensities (c) for the 200~nm
thick Ga$_{0.924}$Mn$_{0.076}$As sample before (open circles) and
after (open squares) annealing as a function of mode index $n$.  (a)
the solid lines show the linear fitting for the experimental data. (c)
The dotted line connects the theoretical result for the normalized
peak intensities. Note that this calculation was done {\em with no
fitting parameters} (see text). Odd modes are not observed
experimentally, as their amplitudes are much smaller compared to those
of the even modes.}
\end{figure}

Fig.~\ref{fig0} shows typical examples of the observed spectra. The
main trends and qualitative features of these data can be effectively
conveyed if the position and linewidth of the resonance features are
plotted as a function of the {\em mode number}.  We label the largest
resonance at the highest field (the uniform FMR mode) as $n=0$, and
the SWR modes as
$ n=1,2,3...$, where the increasing mode index
corresponds to decreasing resonance field. The results of this
procedure are shown in Fig.~\ref{respositions}. The distance between
SWR modes increases for thinner samples.  This is in accordance with
the simple picture, that these resonances are standing waves of
magnetization trapped between the two interfaces of a uniform thin
film, and that the level spacing between successive standing wave
modes increases with decreasing geometrical size of the 'resonator
cavity'. However, from this simplistic picture also follows that the
resonances must depend quadratically on the mode index, in accordance
with Kittel's equation, (\ref{eq:kittel}). This expectation fails
spectacularly in our ${\rm Ga}_{1-x}{\rm Mn}_x{\rm As}$ thin films.
As one can see in Fig.~\ref{respositions} for the 200 nm thick
as-grown and annealed samples, in spite of the fact that we are able
to see as many as 7 spin-wave resonances, the SWR mode positions as a
function of the mode number do not follow the expected quadratic
law. Instead, the resonance fields $H_n$ exhibit a {\it linear
dependence} on the mode index $n$.

We can also compare the intensity of the resonances with Kittel's
predictions. However, to do that we have to keep in mind that the FMR
measurement is done with a field modulation lock-in technique, which
measures the derivative of the absorption as a function of the applied
magnetic field. The observed peak-to peak height $\Delta I_n$ and
peak-to-peak width $\Delta H_n$ of the $n$-th resonance are related to
its intensity $I_n$ as\cite{intensity}
\begin{equation}
I_n \sim \Delta I_n\;\Delta H_n^2
\end{equation}
It is rather striking that the linewidth behaves qualitatively
differently for annealed and unannealed samples: For the as-grown
film the linewidth is independent of $n$ and its value is $\Delta
H_n \sim 200 ~\mbox{Oe}$. For the annealed samples, on the other
hand, $\Delta H_n$ decreases with $n$ (see
Fig.~\ref{respositions}.b), which is a typical behavior of
metallic thin films \cite{typical}. This change to more
conventional behavior in the resonance linewidth corroborates the
observation that the annealed samples are more metallic than the
as-grown samples: resistivity decreases upon annealing, consistent
with an increase in the density of mobile charge carriers
\cite{yu02}. On the other hand, the peak intensities {\em remain}
anomalous despite the annealing. The extracted intensities $I_n$
are shown in Fig.~\ref{respositions}.c. They follow a similar law
for both the annealed and unannealed samples, and the decrease of
the successive intensities is much slower than the $I_n\propto
1/(n+1)^2$ dependence predicted by Kittel.

Clearly, based on the qualitative and quantitative experimental
observations made above, the ${\rm Ga}_{1-x}{\rm Mn}_x{\rm As}$
films do not behave as homogeneous ferromagnetic thin films. The
first attempt to explain this behavior is due to Sasaki {\it
et~al.}\cite{thesis,sasaki03} who, inspired by the work of
Portis\cite{portis63}, suggested a picture where the magnetization
inside the ferromagnetic thin film is not homogeneous in the
direction perpendicular to the plane of the film. Although such
inhomogeneity could explain qualitatively the linear dependence of
successive SWR positions on $n$, the gradient of composition
required to fit the data is unrealistically large, and recent
neutron reflectivity measurements \cite{suzanne} do not support
this degree of variation of magnetization across the sample. An
alternative model is therefore needed for comparing experiment and
theory. We present such a theoretical analysis in the following
section, which will retain in spirit the approach of
Refs.~\onlinecite{thesis,sasaki03}  in that we also propose
inhomogeneity in magnetic parameters along the thickness of the
sample as the main reason behind the observed anomalous behavior
of the SWR positions. However, in our theoretical framework -- in
addition to a gradient of magnetization -- we also allow for a
variation in magnetic anisotropy and spin stiffness. One of the
attractive features of our approach is that -- once the anisotropy
profile has been determined to fit the resonance positions -- it
allows for a {\em parameter-free evaluation} of the normalized
peak intensities (see the points connected by dashed lines in Fig.~
\ref{respositions}.c. ). The results are in excellent agreement
with those extracted from the experiment for {\em both} as-grown
and annealed samples\cite{figure3c}. This agreement in particular
gives us confidence that the theoretical model and numerical
analysis which we present below gives an essentially accurate
description of the FMR and SWR experiments in ${\rm Ga}_{1-x}{\rm
Mn}_x{\rm As}$.

\section{Theoretical approach}
\label{theory}

\subsection{Semiclassical spin wave equations}

It is well accepted that the magnetism of the ${\rm Ga}_{1-x}{\rm
Mn}_x{\rm As}$ is due to indirect interaction between Mn spins
mediated by
holes~\cite{dietl00,jungwirth02,berciu01,zarand02}. Although this
system contains considerable positional disorder -- which, combined
with spin-orbit effects, may lead to non-collinear ground states
\cite{schliemann02,zarand02} -- it appears that on larger length
scales, relevant for the long wavelength collective modes of the
ferromagnetic order parameter, these DMS materials behave as
conventional ferromagnets ~\cite{domains}.  In the following we will
therefore neglect many of the complications which are not relevant
when dealing with spin waves, and we shall employ the usual
semi-classical equations of motion to study spin wave excitations in
the absence of damping~\cite{landau35}.

The first step to derive the semiclassical equations of motion is to
construct the free energy functional for the magnetization
$\vec{\mathbf{M}}({\bf r})$. For temperatures below the transition
temperature $T_C$ this can be expressed as
\begin{widetext}
\begin{equation}
F( \vec{\mathbf{M}})=\int_V d^3r\, \frac{A}{M^2}|\nabla
\vec{\mathbf{M}}|^2+\int_V d^3r\, \frac{K}{M^2}|\vec{\mathbf{M}}\cdot
\vec{u}\,|^2-\int_V d^3r\, (\vec{\mathbf{H}}_{ext}-2\pi
\vec{\mathbf{M}})\cdot\vec{\mathbf{M}}\;. \label{freeEnergy}
\end{equation}
\end{widetext}
In Eq.~(\ref{freeEnergy}), $M=|\vec{\mathbf{M}}|$ denotes the
magnetization of the sample at temperature $T$, and $H_{ext} ||
\hat{z}$ is the applied $dc$ magnetic field. Every integration runs
over the volume $V$ of the film. Note that the free energy functional
in Eq.~(\ref{freeEnergy}) only describes {\em transverse} fluctuations
of the magnetization, and the longitudinal fluctuations are assumed to
be negligible, $M=|\vec{\mathbf{M}}|=constant$.

The first term in Eq.~(\ref{freeEnergy}) is an exchange free energy,
with $A({\bf r})$ denoting the spin wave stiffness, while the second
term represents - depending on the sign of $K({\bf r})$ - a uniaxial
or an in-plane anisotropy energy with respect to direction
$\vec{u}$. The primary source of the anisotropy $K$ is the strain
field due to the lattice mismatch between the GaAs substrate and the
ferromagnetic ${\rm Ga}_{1-x}{\rm Mn}_x{\rm As}$
film\cite{abolfath01a}. Since the ${\rm Ga}_{1-x}{\rm Mn}_x{\rm As}$
films discussed throughout this paper were deposited on ${\rm Ga As}$,
they have an in-plane easy axis which corresponds to $\vec{u}=(0,0,1)$ and
positive $K$. We have not included other cubic anisotropy terms in
Eq.~(\ref{freeEnergy}) because at temperatures they are small compared
to $|K|$. Finally, the last term in Eq.~\ref{freeEnergy} is the Zeeman
free energy together with the demagnetization energy.  Note that all
coefficients in Eq.~(\ref{freeEnergy}) depend on temperature $T$, and
on {\em spatial coordinates}.

Given the material specifics detailed above, the semiclassical
equations of motion can be written as follows:
\begin{equation}\label{eqmot6}
\frac{{d} \vec{\mathbf{M}}(\vec{r}, t)}{{
d}t}=-|\gamma|\vec{\mathbf{M}}(\vec{r},
t)\times\vec{\mathbf{H}}_{tot}(\vec{r}, t).
\end{equation}
Here $\gamma$ is the gyromagnetic ratio, which we approximate by that
of the Mn core spins, $\gamma\approx \gamma_e=2
\mu_B/\hbar$. $\vec{\mathbf{H}}_{tot}(\vec{r},t)$ denotes the total
effective magnetic field, which can be obtained by taking the
functional derivative of the free energy functional,
Eq.~(\ref{freeEnergy}), with respect to the magnetization, $
\vec{\mathbf{H}}_{\rm tot}(\vec{r},t)=-{\delta
F(\{\vec{\mathbf{M}}\})}/{\delta \vec{\mathbf{M}}(\vec{r},t)}$. The
result at $T=0$~K temperature is
\begin{equation}
\vec{\mathbf{H}}_{\rm tot}(\vec{r},t)=-\Lambda \nabla^2
\vec{\mathbf{M}}(\vec{r},t)-\frac{2K}{M}\hat{z}+\vec{\mathbf{H}}_{ext}-4\pi
\vec{\mathbf{M}}\;, \label{Heff}
\end{equation}
where we assumed that the exchange constant $ \Lambda = \frac{2A}{M^2}
$ does not depend on the spatial coordinate.  The first term in
Eq.~\ref{Heff} is usually called the exchange field, while the second
term defines the uniaxial anisotropy field,
$$
\vec{H}_a=-\frac{2K}{M}\hat{z}\;.
$$

To compute the spin wave spectrum for a field parallel to $\hat z$ we
expand the magnetization around its equilibrium value and orientation
$M \hat z$ and, as we mentioned before, we allow for transverse
deviations only:
$\vec{\mathbf{M}}=M\hat{z}+\vec{\mathbf{m}}(\vec{r},t)$. Assuming
periodic time dependence and plane wave character of the {\em r.f.}
magnetization in the x and y directions,
$\vec{\mathbf{m}}(\vec{r},t)=\left[m_x(z)\hat{x}+m_y(z)\hat{y}\,\right]e^{i\omega
t}e^{ik_xx+ik_yy} $, one obtains the following equation of motion for
\mbox{$m^+(z)\equiv m_x(z)+im_y(z)$} in the long wavelength limit
$k_x,k_y\to0$:
\begin{widetext}
\begin{equation}
\{-\Lambda M\frac{d^2 }{d z^2}+4\pi M(z)-H_a(z)-\Lambda\frac{d^2 M(z)}{d
z^2}\}\,m^+(z)=(H_{ext}-\frac{\omega_r}{\gamma})\,m^+(z)\;, \label{eqmot5}
\end{equation}
\end{widetext}
where $\omega_r$ is the frequency of the microwave radiation.  In
course of the derivation of Eq.~\ref{eqmot5} we assumed that only the
magnetic anisotropy and the magnetization vary along the Z direction
(growth direction) ($H_a=H_a(z)$ and $M=M(z)$), but (as already
mentioned after Eq. (5)) that $\Lambda(z)=\Lambda$ is constant. While
these assumptions need not be valid in general, we will show below
that the experimental data on our DMS films can be well explained by
the presence of a
 spatially-dependent anisotropy field
\cite{goennenwein03}. As mentioned before, a spatially varying
magnetization can also reproduce qualitatively the main features of
the data\cite{sasaki02}, but if one allows only $M(z)$ to vary, we now
find by numerically solving Eq.~(\ref{eqmot5}) an unrealistic solution
that cannot reproduce quantitatively the experimental SWR
spectrum.  There is yet another possibility, that the exchange
constant is itself spatially dependent, $\Lambda = \Lambda (z)$. Given
that $ \Lambda $ depends on the carrier concentration, a scenario with
a $z$-dependent exchange constant would, in principle, be consistent
with recent experiments of Koeder and coworkers\cite{koeder}. While we
cannot exclude this possibility, we expect that this would give a
spectrum similar to that of the spatially non-uniform magnetization,
and therefore a very large spatial dependence of the carrier
concentration would be needed to explain the experimental data. We
therefore classify this possibility, together with that of
spatially-dependent magnetization, as {\em sub-dominant} mechanisms,
which would at most play a secondary role in explaining the features
of our resonance experiments. The versatility of our numerical
analysis is clearly evident in evaluating the different scenarios
discussed above: our numerical scheme allows us to check all these
scenarios against our resonance absorption data and the constrains set
by other experiments, and select the most viable one for explaining
our FMR and SWR measurements.

 Equation~(\ref{eqmot5}) has the form of a one-dimensional
Schr\"{o}dinger equation for an electron in a quantum well. If we
compare the coefficients in Eq.\ref{eqmot5} with the Schr\"{o}dinger
equation, we can see that $\hbar^2/(2\Lambda M(z))$ is the analog of
the electron mass (in our case it may depend on the position $z$ if
$M$ varies with position) and $4\pi M(z)-H_a(z)-\Lambda\frac{d^2
M(z)}{d z^2}$ is the analog of the potential energy
$V(z)$. Equation~(\ref{eqmot5}) can then be formally written as
\begin{equation}
[\frac{p^2}{2m}+V(z)]m^+_n(z)=E_nm^+_n(z), \label{eq:schrodinger}
\end{equation}
where $p=-i\hbar\frac{d}{dz}$ is the momentum operator and
$E_n=H_{ext}-\frac{\omega}{\gamma}$ is the $n$-th eigenvalue. Having
solved this Schr\"odinger equation, we can compute the intensities
$I_n$ of each mode as\cite{sparks70c}
\begin{equation}\label{intensity}
I_n \propto \frac{|\int_{0}^{L} m^+_n(z) dz|^2}{\int_{0}^{L} |m^+_n(z)|^2 dz},
\end{equation}
where the integration runs over the ferromagnetic film thickness.

\subsection{Boundary conditions and intensity of SWR}

To solve Eq.~(\ref{eqmot5}), it is crucial to establish the boundary
conditions for SWR.  While different boundary conditions do not affect
the \textit{positions} of the resonances in an essential way, the
\textit{intensity} of the SWR peaks strongly depends on them. It is
therefore important to choose the appropriate boundary conditions
(BCs).

The question of boundary conditions has been debated for a rather long
time in the literature and, to our knowledge, no general agreement has
been reached. Different authors~\cite{hoekstra77,sparks69,soohoo63}
treat the BC problem in different ways, and assume different boundary
conditions appropriate to the specific sample which they are studying.

Equation~(\ref{eqmot5}) must be, in general, solved together with
macroscopic Maxwell equations. Ament and Rado~\cite{Ament55} argued
that in the absence of any magnetic anisotropy the spin wave solutions
should satisfy so-called free (or anti-node) boundary conditions
($\frac{dm^+(z)}{dz}=0$ at $z=0$ and $z=L$). On the another hand,
Pincus~\cite{pincus60} pointed out that Ament and Rado have not
treated properly the discontinuity of the space at the surface of the
film. Pincus (and also previously Kittel~\cite{kittel49}) indicated
that at the interface an additional term, proportional to the gradient
of the magnetization at the surface (not included in equation
\ref{eqmot5}), is also allowed by symmetry, and that this term can
lead to the pinning of spin waves at the boundaries. Furthermore,
Pincus observed that free BCs can never be appropriate when there is
surface anisotropy. He also showed that an antiferromagnetic oxide
layer on the surface of a film can give rise to a surface anisotropy
which pins the spins at the end-points ($m^+_n(z=0)=0$ and
$m^+_n(z=L)=0$). Finally, Pincus and Kittel have shown that it is more
appropriate to use the so-called \textit{dynamic boundary
conditions},\cite{pincus60,kittel49} which for the lowest lying modes
effectively reduce to pinned boundary conditions.

\begin{figure}
\includegraphics[width=7cm]{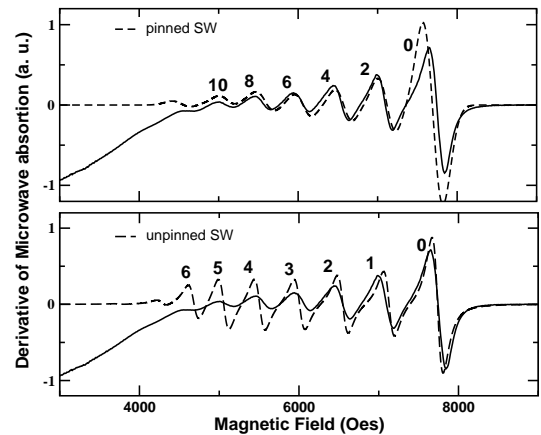} \vskip0.3cm
\caption{\label{bc_dependence} The derivative of the absorption of the
SWR as a function of applied dc magnetic field for an as-grown 200 nm
\GaMnAs{} film. The solid lines represent the experimental data, and
the dashed lines show our theoretical results for the magnetic
anisotropy profile of Eq.~(\protect\ref{pro1}). The upper/lower panels show
the results of our computations for pinned/unpinned boundary
conditions.}
\end{figure}

We believe that in the case of ${\rm Ga}_{1-x}{\rm Mn}_x{\rm As}$
films one must also use {\em pinned} boundary conditions. This is
because, first, there is a strong strain field present in these films
that generates a large anisotropy field, and second, we expect a
strong surface-induced anisotropy in the vicinity of the surface due
to the strong spin-orbit coupling in ${\rm Ga}_{1-x}{\rm Mn}_x{\rm
As}$, which would also pin the spin waves\cite{ujsaghy}. Fortunately,
our numerical analysis allows us to try different boundary
conditions. We were unable to obtain a good fit to the peak
intensities
when we used unpinned or partially unpinned boundary
conditions. Pinned boundary conditions, on the other hand, gave very
good agreement with the measured SWR spectra. This is illustrated in
Fig.~\ref{bc_dependence}, where we show our best fits with pinned and
unpinned boundary conditions for the SWR spectra for an as-grown 200
nm sample at $T=4 ~{\rm K}$.

\begin{figure}
\includegraphics[width=9.0cm]{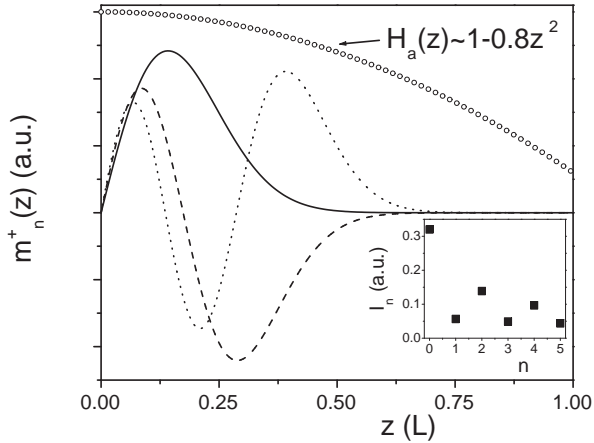}
\caption{Ilustration of the functional form of three spin wave modes
with $n=0$, 1 and 2 (solid, dashed, and dotted lines) for the as-grown
$L=200 {\;\rm nm}$ sample at $T=4{\rm K}$.  We used an exchange
constant $\Lambda = 3000 {\;\rm nm}^2$ and $\alpha=0.8$. The rescaled
anisotropy profile is marked by open circles. Distances are measured
in units of $L$. In the inset we show the calculated intensities $I_n$
of the modes in the main figure \label{m_of_z}.}
\end{figure}

Typical solutions for pinned boundary conditions and the corresponding
intensities are shown in Fig.~\ref{m_of_z}. The intensity of odd modes
is suppressed, because of the cancellation of the regions with $m^+>0$
and $m^+<0$, respectively.  In fact, in Fig.~\ref{bc_dependence} the
odd modes are barely visible, and only peaks associated with the {\em
even modes} can be observed.

The low energy spin (standing) waves tend to be localized around the
region with higher magnetic anisotropy. Qualitatively speaking, this
is because we have the {\em negative} of the spatially-dependent
uniaxial anisotropy acting as the effective trapping potential for the
modes, and therefore it is easier to create spin waves where the
anisotropy is large. For smaller values of $\Lambda$, spin waves tend
to be more localized around the large anisotropy points. On the other
hand,  spin waves with higher energies
will be extended over the whole sample thickness, and their energy
will not depend linearly on $n$.

\section{Numerical Results and Discussion}
\label{results}

We solved Eq.~(\ref{eqmot5}) numerically, and varied the exchange
parameter $\Lambda$ and the profiles of the anisotropy field $H_a(z)$
(or the magnetization $M(z)$) to obtain a best fit to the experimental
spectra. From these calculations we concluded that to explain a linear
variation of the resonance fields, $H_n\sim n$, the changes of
$H_a(z)$ (or $M(z)$) must be substantial across the whole depth of the
sample. For profiles where the variation of $H_a(z)$ (or $M(z)$) is
only near the surface, the resonance positions show always a quadratic
behavior, $H_n\sim n^2$. This can be easily understood from
Eq.~(\ref{eq:schrodinger}), since in this case the potential $V(z)$ is
similar to that of a rectangular quantum well and the eigen-modes are
therefore quadratically separated.

As discussed in the previous section, the spatial variation of the
magnetization and the exchange constant are sub-dominant
processes. On the other hand, we were able to obtain quantitative
agreement with the experimental data allowing the magnetic anisotropy
field $H_a=H_a(z)$ to vary across the sample while keeping the
magnetization constant, $M(z)=const$. Therefore, in the discussion
that follows, we shall mostly focus to the case of a $z$-dependent
anisotropy field, $H_a=H_a(z)$, and assume $M=const$.

The computed SWR spectrum depends on the specific shape of
the anisotropy profile $H_a(z)$. We found, in particular, that a
linear dependence of $H_a$ on $z$, $H_a(z)\sim z$, is clearly in
disagreement with our experimental data.
%
However, we could obtain an excellent fit by assuming a {\em
quadratic} dependence on $z$, $H_a(z)\sim z^2$. More specifically,
both of the following profiles fit the experimental data rather well:
\begin{eqnarray}
H_a(z)&=&H_a \exp{(-\alpha z^2/L^2)} \label{pro1}\\ &=&H_a(1-\alpha
\,z^2/L^2),\label{pro2} \,\,\,\,\;,
\end{eqnarray}
where $L$ is the thickness of the film, $\alpha\sim 0.75$ is a
dimensionless fitting parameter, and the maximum value of the
anisotropy field, $H_a(0)$, has been extracted from the analysis of
the main FMR resonance~\cite{liu03}. Since we can obtain the value of
the saturation magnetization $M$ from accurate SQUID measurements, we
end up with only two fitting parameters, $\alpha$ and the stiffness
$A$ (or equivalently, the exchange constant $\Lambda$).

The origin of the linear behavior on $n$ can be easily understood from
the formal analogy with the Schr\"odinger equation
(Eq.~\ref{eq:schrodinger}) and the energy spectrum of the harmonic
oscillator: as long as the difference between the resonance field
$H_n$ and the maximum value of the anisotropy field $|H_a(0)|-H_n$ is
small compared to $\frac{\omega_r}{\gamma}$, we expect the
corresponding wave function $m^+(z)$ to be well approximated by the
Hermite functions and to have a linear behavior $\sim n$. This
condition is well satisfied for the experimentally observed spin wave
resonance fields.

\begin{figure}[b]
\includegraphics[width=6.5cm,clip]{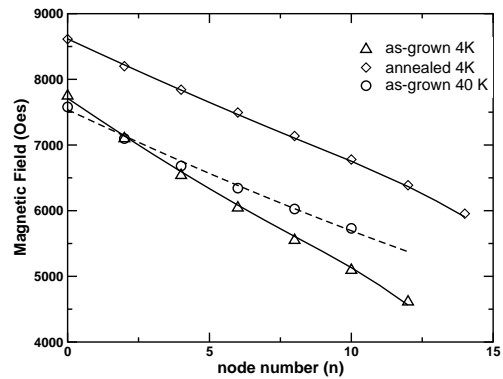}
\caption{\label{fig3} Comparison between theoretical (solid lines) and
experimental (symbols) data for the 200 nm sample. Two different
temperatures for the as-grown sample and one temperature for the
annealed sample are considered. The saturation magnetization data from
SQUID measurements is used in the theoretical approach.}
\end{figure}

In order to study the anisotropy profile, we focused on the 200 nm
sample, because this sample exhibited the largest number of resonance
peaks.  First, we chose our parameters to fit the 4 K data
(Fig.~\ref{fig3}): Having obtained the magnetization $M=
17.9\;\mbox{emu/cm$^3$}$ from SQUID experiments, and the magnetic
anisotropy field $H_a= 4.400$~Oe from the angular dependence of the
main FMR resonance line, we adjusted $\Lambda=3000\pm 300 {\;\rm
mn}^2$ and $\alpha=0.75$ to obtain the best fit which corresponds to a spin stiffness {\bf A = 0.4 meV/$\bf \AA$}. For the spectrum
measured at $40 {\;\rm K}$ in the as-grown sample we obtained a very
good fit with the {\em same} exchange constant $\Lambda$ and $\alpha$,
but replacing $H_a(0)$ by the measured $40 \;{\rm K}$ anisotropy
$H^{40 K}_a(0)=4.200 \;\mbox{Oe}$ and magnetization $M^{40 K}=9.4\;
\mbox{emu/cm$^3$}$.  In both cases the SWR spectra (and thus the
resonance fields $H_n$) exhibit a {\em quasi-linear} dependence on
$n$.

Annealing of \GaMnAs{} has various - presently not entirely understood
- effects. First, the saturation magnetization increases upon
annealing. Furthermore, Rutherford backscattering experiments
suggest\cite{Rutherford} that the primary process resulting from
annealing is the diffusion of interstitial Mn ions out of the film.
Since interstitial Mn takes away carriers from the hole band,
annealing results probably in an {\em increase of the carrier
density}, and the resulting carrier density is most likely {\em
inhomogeneous}. Since both the anisotropy and the exchange energy are
sensitive to the carrier density, annealing is expected to change the
value of both $H_a$ and $\Lambda$,\cite{abolfath01} and it may also
change their profiles.

Consequently, it is natural to expect that the SWR spectra observed on
annealed specimens will require fitting parameters that are different
from the ones used for the as-grown sample. Nevertheless, we were able
to obtain a good fit to the experimental SWR by keeping the profile
unchanged ($\alpha=0.75$), adjusting the anisotropy field $H_a^{\rm
anneal}(0)=5300$~Oe and using $\Lambda=1225 \;{\rm
nm}^2$. Surprisingly, this value of exchange constant $\Lambda$
corresponds to the same value of the stiffness parameter $A$ as in the
as-grown sample.

\section{Conclusions}
We have analyzed the experimental FMR and SWR spectra of thin ${\rm
Ga}_{1-x}{\rm Mn}_x{\rm As}$ films grown on GaAs substrates by low
temperature MBE technique. We compared the experimentally observed
resonance positions and intensities with theoretical calculations. The
experimental results are clearly inconsistent with the assumption of a
homogeneous ${\rm Ga}_{1-x}{\rm Mn}_x{\rm As}$ film. After solving the
semiclassical equations of motion and using independent experimental
data for the magnetic anisotropy and magnetization, we have shown that
a nearly quadratic $z$-dependence of the uniaxial anisotropy,
$H_a(z)-H_a(0) \sim z^2$ could adequately explain the position and
intensity of the experimental data. An inhomogeneous magnetization
could also lead to similar effects. However, the spatial variation and
the magnitude of magnetization needed to explain the experimental data
is incompatible with the
experiments. Spatial variation of the
magnetization, as observed in
Ref.~\onlinecite{suzanne} play a secondary role in
the SWR spectra. From our numerical analysis we also concluded that
the observed spin waves are pinned at the surfaces of the film. As a
consequence of pinning, the intensity of $n={\rm odd}$ spin wave modes
is suppressed and the experimentally observed spectra consist of even
modes only.

Goennenwein~{\it et~al.} performed similar SWR measurements and also
found spin wave spectra that are incompatible with having homogeneous
${\rm Ga}_{1-x}{\rm Mn}_x{\rm As}$ films\cite{goennenwein03}. They,
however, find that $H_n\propto n^{2/3}$, consistent with a linear
$z$-dependence of the anisotropy field, $H_a(z)-H_a(0) \sim z$. This
is in qualitative contrast with our findings for the resonance fields
which scale as $H_n\sim n$. Furthermore, Goennenwein {\em et al.}
appear to have implicitly assumed free boundary conditions, which
implies that both even and odd modes are observable. In contrast, we
argue that \textit{pinned} boundary conditions are more adequate for
our samples, and we find that, again, only even modes can be
observed.

There may be several mechanisms that produce inhomogeneous magnetic
parameters. The uniaxial anisotropy is primarily due to uniaxial
strain field in the film that develops due to the lattice mismatch
between the substrate and the ferromagnetic
film\cite{ohno96,liu03,abolfath01}. In principle, an inhomogeneous
strain field could therefore produce an inhomogeneous anisotropy field
$H_a$. However, in MBE grown samples the strain usually relaxes
suddenly, when the sample reaches a critical thickness, where
dislocations start to form and relax the strain. In these
non-equilibrium MBE grown samples, however, there seems to be no
strain field relaxation at all through dislocation formation: Even for
micron thick samples the measured in-plane lattice constant of the
film is the same as that of the substrate.

One can also obtain inhomogeneous magnetic properties by assuming an
inhomogeneous {\em hole concentration}. Indeed, annealing experiments
support the notion that the dominant compensation effect is due to
interstitial Mn ions, which act as double donors and may also
compensate the spin of substitutional Mn. These interstitial Mn ions
can diffuse out of the sample during the growth process, which takes
usually a few hours and takes place at the same temperature as the
annealing. As a result, it is quite possible that the concentration of
charge carriers (related to that of Mn interstitials) varies across
the film. Since the exchange anisotropy and exchange constants are
both related to the carrier density\cite{abolfath01a}, this can serve
as a mechanism to produce a $z$-dependent anisotropy field. Several
recent experiments provide firm support for an inhomogeneous hole
concentration: Koeder and coworkers\cite{koeder} find indications of
gradients in both the carrier concentration and Curie temperature of
epitaxial ${\rm Ga}_{1-x}{\rm Mn}_x{\rm As}$ films. The presence of
such gradients are also in accord with recent observations of
interstitial Mn diffusion \cite{yu02,gallagher}. The notion that Mn
diffusion is a key process for annealing-induced enhancement of
magnetism has recently been backed up by annealing experiments on
samples {\em capped}\cite{shiffer-cap} with a few monolayers of
undoped GaAs, which would block the interstitial Mn from diffusing out
of the sample. Under these circumstances the annealing process gave no
significant increase in the Curie temperature. We believe that these
experimental results, together with our resonance measurements and the
theoretical arguments given above provide solid support for our
calculation that a gradient in the magnetic parameters must be present
in ${\rm Ga}_{1-x}{\rm Mn}_x{\rm As}$.

We also studied the linewidth of the observed resonances. For the
as-grown samples the SWR linewidths do not depend on the mode
number. This behavior hints that the relaxation is due to spin-orbit
coupling.\cite{zarand02} Indeed, the measured (rather large) resonance
width is compatible with the presence of a relatively large random
anisotropy.

For the annealed samples, on the other hand, the linewidth decreases
with increasing mode number, which is characteristic to eddy current
relaxation in metallic samples\cite{typical}. This behavior is
consistent with the results of annealing experiments, since for higher
carrier densities the random anisotropy effects become less important,
and at the same time the sample becomes more metallic. It also
underscores the fact that there is a significant physical difference
between as-grown and annealed samples, going beyond quantitative
changes in the values of saturation magnetization and the Curie
temperature.

Since the variation of elastic and/or magnetic properties across the
\GaMnAs{} film can have important consequences in its future
spintronics applications, a more detailed experimental and theoretical
analysis is necessary to understand and control the magnetic
properties of these thin ferromagnetic layers. One way to obtain more
information about the effects of the surface anisotropy is to perform
FMR measurements in symmetric three-layer structures \mbox{GaAs/${\rm
Ga}_{1-x}{\rm Mn}_x{\rm As}$/GaAs} and compare them with the
previously obtained results. A systematic study of the thickness and
annealing time dependence of similarly grown samples would be also
important to understand the origin of the observed gradient of
composition.

\begin{section}{Acknowledgment}
We would like to thanks Prof. John B. Ketterson for important
comments. This research was supported by the National Science
Foundation under NSF-NIRT award DMR 02-10519, by the U.S. Department
of Energy, Basic Energy Sciences, under Contract No. W-7405-ENG-36, by
the Alfred P. Sloand Foundation (B.J.), and by the NSF and Hungarian
Grants No. OTKA F030041, and T038162.
G.Z. is a Bolyai Fellow of the Hungarian Academy of Sciences.
\end{section}

\bibliography{paper}

\begin{thebibliography}{43}
\expandafter\ifx\csname natexlab\endcsname\relax\def\natexlab#1{#1}\fi
\expandafter\ifx\csname bibnamefont\endcsname\relax
  \def\bibnamefont#1{#1}\fi
\expandafter\ifx\csname bibfnamefont\endcsname\relax
  \def\bibfnamefont#1{#1}\fi
\expandafter\ifx\csname citenamefont\endcsname\relax
  \def\citenamefont#1{#1}\fi
\expandafter\ifx\csname url\endcsname\relax
  \def\url#1{\texttt{#1}}\fi
\expandafter\ifx\csname urlprefix\endcsname\relax\def\urlprefix{URL }\fi
\providecommand{\bibinfo}[2]{#2}
\providecommand{\eprint}[2][]{\url{#2}}

\bibitem[{\citenamefont{Ohno}(1998)}]{ohno98}
\bibinfo{author}{\bibfnamefont{H.}~\bibnamefont{Ohno}},
  \bibinfo{journal}{Science} \textbf{\bibinfo{volume}{281}},
  \bibinfo{pages}{951} (\bibinfo{year}{1998}).

\bibitem[{\citenamefont{Furdyna et~al.}(2000)\citenamefont{Furdyna, Schiffer,
  Potashnik, and Liu}}]{furdyna00}
\bibinfo{author}{\bibfnamefont{J.~K.} \bibnamefont{Furdyna}},
  \bibinfo{author}{\bibfnamefont{S.}~\bibnamefont{Schiffer},
  \bibfnamefont{P.and~Sasaki}},
  \bibinfo{author}{\bibfnamefont{J.}~\bibnamefont{Potashnik}},
  \bibnamefont{and} \bibinfo{author}{\bibfnamefont{X.~Y.} \bibnamefont{Liu}},
  \emph{\bibinfo{title}{Optical Properties od Semiconductor Nanostructures}},
  vol.~\bibinfo{volume}{81} (\bibinfo{publisher}{NATO Science Series},
  \bibinfo{year}{2000}).

\bibitem[{\citenamefont{Ohno}(2002)}]{ohno02}
\bibinfo{author}{\bibfnamefont{H.}~\bibnamefont{Ohno}}, \bibinfo{journal}{J.
  Magn. Magn. Mater.} \textbf{\bibinfo{volume}{242}}, \bibinfo{pages}{105}
  (\bibinfo{year}{2002}).

\bibitem[{\citenamefont{Farle}(1998)}]{farle98}
\bibinfo{author}{\bibfnamefont{M.}~\bibnamefont{Farle}}, \bibinfo{journal}{Rep.
  Prog. Phys.} \textbf{\bibinfo{volume}{61}}, \bibinfo{pages}{755}
  (\bibinfo{year}{1998}).

\bibitem[{\citenamefont{Liu et~al.}(2003)\citenamefont{Liu, Sasaki, and
  Furdyna}}]{liu03}
\bibinfo{author}{\bibfnamefont{X.}~\bibnamefont{Liu}},
  \bibinfo{author}{\bibfnamefont{Y.}~\bibnamefont{Sasaki}}, \bibnamefont{and}
  \bibinfo{author}{\bibfnamefont{J.~K.} \bibnamefont{Furdyna}},
  \bibinfo{journal}{Phys. Rev. B} \textbf{\bibinfo{volume}{67}},
  \bibinfo{pages}{205207} (\bibinfo{year}{2003}).

\bibitem[{\citenamefont{Fedorych et~al.}(2002)\citenamefont{Fedorych,
  Byszewska, Wilamowski, Potemski, and Sadowski}}]{fedorych02}
\bibinfo{author}{\bibfnamefont{O.}~\bibnamefont{Fedorych}},
  \bibinfo{author}{\bibfnamefont{M.}~\bibnamefont{Byszewska}},
  \bibinfo{author}{\bibfnamefont{Z.}~\bibnamefont{Wilamowski}},
  \bibinfo{author}{\bibfnamefont{M.}~\bibnamefont{Potemski}}, \bibnamefont{and}
  \bibinfo{author}{\bibfnamefont{J.}~\bibnamefont{Sadowski}},
  \bibinfo{journal}{Acta Phys. Pol. A} \textbf{\bibinfo{volume}{102}},
  \bibinfo{pages}{617} (\bibinfo{year}{2002}).

\bibitem[{\citenamefont{Kittel}(1958)}]{kittel58}
\bibinfo{author}{\bibfnamefont{C.}~\bibnamefont{Kittel}},
  \bibinfo{journal}{Phys. Rev} \textbf{\bibinfo{volume}{110}},
  \bibinfo{pages}{1295} (\bibinfo{year}{1958}).

\bibitem[{spi()}]{spinstiff}
\bibinfo{note}{The stiffness $D$ is related to the constant $A$ in Section
  \ref{theory} as $D=2 A/M$ where $M$ is the magnetization.}

\bibitem[{\citenamefont{Goennenwein~{\it et al.}}(2003)}]{goennenwein03}
\bibinfo{author}{\bibfnamefont{S.~T.~B.} \bibnamefont{Goennenwein~{\it et
  al.}}}, \bibinfo{journal}{Appl. Phys. Lett.} \textbf{\bibinfo{volume}{82}},
  \bibinfo{pages}{730} (\bibinfo{year}{2003}).

\bibitem[{\citenamefont{Ohno et~al.}(1996)\citenamefont{Ohno, Shen, Matsukura,
  Oiwa, Endo, Katsumoto, and Iye}}]{ohno96}
\bibinfo{author}{\bibfnamefont{H.}~\bibnamefont{Ohno}},
  \bibinfo{author}{\bibfnamefont{A.}~\bibnamefont{Shen}},
  \bibinfo{author}{\bibfnamefont{F.}~\bibnamefont{Matsukura}},
  \bibinfo{author}{\bibfnamefont{A.}~\bibnamefont{Oiwa}},
  \bibinfo{author}{\bibfnamefont{A.}~\bibnamefont{Endo}},
  \bibinfo{author}{\bibfnamefont{S.}~\bibnamefont{Katsumoto}},
  \bibnamefont{and} \bibinfo{author}{\bibfnamefont{Y.}~\bibnamefont{Iye}},
  \bibinfo{journal}{Appl. Phys. Lett.} \textbf{\bibinfo{volume}{69}},
  \bibinfo{pages}{363} (\bibinfo{year}{1996}).

\bibitem[{\citenamefont{Abolfath
  et~al.}(2001{\natexlab{a}})\citenamefont{Abolfath, Jungwirth, and
  MacDonald}}]{abolfath01}
\bibinfo{author}{\bibfnamefont{M.}~\bibnamefont{Abolfath}},
  \bibinfo{author}{\bibfnamefont{T.}~\bibnamefont{Jungwirth}},
  \bibnamefont{and} \bibinfo{author}{\bibfnamefont{A.~H.}
  \bibnamefont{MacDonald}}, \bibinfo{journal}{Physica E}
  \textbf{\bibinfo{volume}{10}}, \bibinfo{pages}{161}
  (\bibinfo{year}{2001}{\natexlab{a}}).

\bibitem[{\citenamefont{Dietl et~al.}(2001)\citenamefont{Dietl, Konig, and
  MacDonald}}]{dietl01}
\bibinfo{author}{\bibfnamefont{T.}~\bibnamefont{Dietl}},
  \bibinfo{author}{\bibfnamefont{J.}~\bibnamefont{Konig}}, \bibnamefont{and}
  \bibinfo{author}{\bibfnamefont{A.~H.} \bibnamefont{MacDonald}},
  \bibinfo{journal}{Phys. Rev. B} \textbf{\bibinfo{volume}{64}},
  \bibinfo{pages}{241201} (\bibinfo{year}{2001}).

\bibitem[{\citenamefont{Sasaki et~al.}(2002)\citenamefont{Sasaki, Liu, Furdyna,
  Palczewska, Szczytko, and Twardowski}}]{sasaki02}
\bibinfo{author}{\bibfnamefont{Y.}~\bibnamefont{Sasaki}},
  \bibinfo{author}{\bibfnamefont{X.}~\bibnamefont{Liu}},
  \bibinfo{author}{\bibfnamefont{J.~K.} \bibnamefont{Furdyna}},
  \bibinfo{author}{\bibfnamefont{M.}~\bibnamefont{Palczewska}},
  \bibinfo{author}{\bibfnamefont{J.}~\bibnamefont{Szczytko}}, \bibnamefont{and}
  \bibinfo{author}{\bibfnamefont{A.}~\bibnamefont{Twardowski}},
  \bibinfo{journal}{J. Appl. Phys.} \textbf{\bibinfo{volume}{91}},
  \bibinfo{pages}{7484} (\bibinfo{year}{2002}).

\bibitem[{\citenamefont{Kaminski and Sarma}(2002)}]{kaminski02}
\bibinfo{author}{\bibfnamefont{A.}~\bibnamefont{Kaminski}} \bibnamefont{and}
  \bibinfo{author}{\bibfnamefont{S.~D.} \bibnamefont{Sarma}},
  \bibinfo{journal}{Phys. Rev. Lett.} \textbf{\bibinfo{volume}{88}},
  \bibinfo{pages}{247202} (\bibinfo{year}{2002}).

\bibitem[{\citenamefont{Berciu and Bhatt}(2001)}]{berciu01}
\bibinfo{author}{\bibfnamefont{M.}~\bibnamefont{Berciu}} \bibnamefont{and}
  \bibinfo{author}{\bibfnamefont{R.~N.} \bibnamefont{Bhatt}},
  \bibinfo{journal}{Phys. Rev. Lett.} \textbf{\bibinfo{volume}{87}},
  \bibinfo{pages}{107203} (\bibinfo{year}{2001}).

\bibitem[{\citenamefont{Fiete et~al.}(2003)\citenamefont{Fiete, Zarand, and
  Damle}}]{fiete03}
\bibinfo{author}{\bibfnamefont{G.~A.} \bibnamefont{Fiete}},
  \bibinfo{author}{\bibfnamefont{G.}~\bibnamefont{Zarand}}, \bibnamefont{and}
  \bibinfo{author}{\bibfnamefont{K.}~\bibnamefont{Damle}},
  \bibinfo{journal}{Phys. Rev. Lett.} \textbf{\bibinfo{volume}{91}},
  \bibinfo{pages}{097202} (\bibinfo{year}{2003}).

\bibitem[{\citenamefont{Sasaki et~al.}(2003)\citenamefont{Sasaki, Liu, and
  Furdyna}}]{sasaki03}
\bibinfo{author}{\bibfnamefont{Y.}~\bibnamefont{Sasaki}},
  \bibinfo{author}{\bibfnamefont{X.}~\bibnamefont{Liu}}, \bibnamefont{and}
  \bibinfo{author}{\bibnamefont{Furdyna}}, \bibinfo{journal}{J. Supercond.}
  \textbf{\bibinfo{volume}{16}}, \bibinfo{pages}{41} (\bibinfo{year}{2003}).

\bibitem[{int()}]{intensity}
\bibinfo{note}{The intensity of a resonance can be modelled as {\protect{$I(H)
  = (I_0/\Delta H) f(H/\Delta H)$ }}where the total integrated intensity is
  {\protect{$I_0$, and $\int^{+\infty}_{-\infty} dx f(x) = 1$}}. Then the
  peak-to-peak change in the {\em derivative} of the intensity is
  {\protect{$\Delta I = |f'(1/2) (I_0/\Delta H^2)$.}}}

\bibitem[{\citenamefont{Rado and Asment}(1955)}]{typical}
\bibinfo{author}{\bibfnamefont{G.}~\bibnamefont{Rado}} \bibnamefont{and}
  \bibinfo{author}{\bibfnamefont{W.}~\bibnamefont{Asment}},
  \bibinfo{journal}{Phys. Rev.} \textbf{\bibinfo{volume}{97}},
  \bibinfo{pages}{1558} (\bibinfo{year}{1955}).

\bibitem[{\citenamefont{Yu et~al.}(2002{\natexlab{a}})\citenamefont{Yu,
  Walukiewicz, Wojtowicz, Kuryliszyn, Liu, Sasaki, and Furdyna}}]{yu02}
\bibinfo{author}{\bibfnamefont{K.~M.} \bibnamefont{Yu}},
  \bibinfo{author}{\bibfnamefont{W.}~\bibnamefont{Walukiewicz}},
  \bibinfo{author}{\bibfnamefont{T.}~\bibnamefont{Wojtowicz}},
  \bibinfo{author}{\bibfnamefont{I.}~\bibnamefont{Kuryliszyn}},
  \bibinfo{author}{\bibfnamefont{X.}~\bibnamefont{Liu}},
  \bibinfo{author}{\bibfnamefont{Y.}~\bibnamefont{Sasaki}}, \bibnamefont{and}
  \bibinfo{author}{\bibfnamefont{J.~K.} \bibnamefont{Furdyna}},
  \bibinfo{journal}{Phys. Rev. B} \textbf{\bibinfo{volume}{65}},
  \bibinfo{pages}{201303} (\bibinfo{year}{2002}{\natexlab{a}}).

\bibitem[{\citenamefont{Sasaki}(2002)}]{thesis}
\bibinfo{author}{\bibfnamefont{Y.}~\bibnamefont{Sasaki}},
  \bibinfo{journal}{P.h.D Thesis University of Notre Dame}
  (\bibinfo{year}{2002}).

\bibitem[{\citenamefont{Portis}(1963)}]{portis63}
\bibinfo{author}{\bibnamefont{Portis}}, \bibinfo{journal}{Appl. Phys. Lett.}
  \textbf{\bibinfo{volume}{2}}, \bibinfo{pages}{69} (\bibinfo{year}{1963}).

\bibitem[{\citenamefont{Rhyne et~al.}()\citenamefont{Rhyne, Kirby, teVelthuis,
  Hoffmann, Wojtowicz, Liu, and Furdyna}}]{suzanne}
\bibinfo{author}{\bibfnamefont{J.}~\bibnamefont{Rhyne}},
  \bibinfo{author}{\bibfnamefont{B.}~\bibnamefont{Kirby}},
  \bibinfo{author}{\bibfnamefont{S.}~\bibnamefont{teVelthuis}},
  \bibinfo{author}{\bibfnamefont{A.}~\bibnamefont{Hoffmann}},
  \bibinfo{author}{\bibfnamefont{T.}~\bibnamefont{Wojtowicz}},
  \bibinfo{author}{\bibfnamefont{X.}~\bibnamefont{Liu}}, \bibnamefont{and}
  \bibinfo{author}{\bibfnamefont{J.}~\bibnamefont{Furdyna}},
  \bibinfo{note}{(unpublished)}.

\bibitem[{fig()}]{figure3c}
\bibinfo{note}{In Fig.3.c the theoretical results, shown as point connected by
  dashed lines, for the normalized mode intensity were calculated within a
  model where the anisotropy has a {\em quadratic} dependence on the distance
  from the interface. For more realistic model of anisotropy variation, see
  Section \ref{results}}.

\bibitem[{\citenamefont{Dietl et~al.}(2000)\citenamefont{Dietl, Ohno,
  Matsukura, Cibert, and Ferrand}}]{dietl00}
\bibinfo{author}{\bibfnamefont{T.}~\bibnamefont{Dietl}},
  \bibinfo{author}{\bibfnamefont{H.}~\bibnamefont{Ohno}},
  \bibinfo{author}{\bibfnamefont{F.}~\bibnamefont{Matsukura}},
  \bibinfo{author}{\bibfnamefont{J.}~\bibnamefont{Cibert}}, \bibnamefont{and}
  \bibinfo{author}{\bibfnamefont{D.}~\bibnamefont{Ferrand}},
  \bibinfo{journal}{Science} \textbf{\bibinfo{volume}{287}},
  \bibinfo{pages}{1019} (\bibinfo{year}{2000}).

\bibitem[{\citenamefont{Jungwirth et~al.}(2002)\citenamefont{Jungwirth, Konig,
  Sinova, Kucera, and MacDonald}}]{jungwirth02}
\bibinfo{author}{\bibfnamefont{T.}~\bibnamefont{Jungwirth}},
  \bibinfo{author}{\bibfnamefont{J.}~\bibnamefont{Konig}},
  \bibinfo{author}{\bibfnamefont{J.}~\bibnamefont{Sinova}},
  \bibinfo{author}{\bibfnamefont{J.}~\bibnamefont{Kucera}}, \bibnamefont{and}
  \bibinfo{author}{\bibfnamefont{A.~H.} \bibnamefont{MacDonald}},
  \bibinfo{journal}{Phys. Rev. B} \textbf{\bibinfo{volume}{66}},
  \bibinfo{pages}{012402} (\bibinfo{year}{2002}).

\bibitem[{\citenamefont{Zarand and Janko}(2002)}]{zarand02}
\bibinfo{author}{\bibfnamefont{G.}~\bibnamefont{Zarand}} \bibnamefont{and}
  \bibinfo{author}{\bibfnamefont{B.}~\bibnamefont{Janko}},
  \bibinfo{journal}{Phys. Rev. Lett.} \textbf{\bibinfo{volume}{89}},
  \bibinfo{pages}{47201} (\bibinfo{year}{2002}).

\bibitem[{\citenamefont{Schliemann and MacDonald}(2002)}]{schliemann02}
\bibinfo{author}{\bibfnamefont{J.}~\bibnamefont{Schliemann}} \bibnamefont{and}
  \bibinfo{author}{\bibfnamefont{A.~H.} \bibnamefont{MacDonald}},
  \bibinfo{journal}{Phys. Rev. Lett.} \textbf{\bibinfo{volume}{88}},
  \bibinfo{pages}{137201} (\bibinfo{year}{2002}).

\bibitem[{\citenamefont{Welp et~al.}(2003)\citenamefont{Welp, Vlasko-Vlasov,
  Liu, Furdyna, and Wojtowicz}}]{domains}
\bibinfo{author}{\bibfnamefont{U.}~\bibnamefont{Welp}},
  \bibinfo{author}{\bibfnamefont{V.~K.} \bibnamefont{Vlasko-Vlasov}},
  \bibinfo{author}{\bibfnamefont{X.}~\bibnamefont{Liu}},
  \bibinfo{author}{\bibfnamefont{J.~F.} \bibnamefont{Furdyna}},
  \bibnamefont{and}
  \bibinfo{author}{\bibfnamefont{T.}~\bibnamefont{Wojtowicz}},
  \bibinfo{journal}{Phys. Rev. Lett.} \textbf{\bibinfo{volume}{90}},
  \bibinfo{pages}{167206} (\bibinfo{year}{2003}).

\bibitem[{\citenamefont{Landau and Lifshitz}(1935)}]{landau35}
\bibinfo{author}{\bibfnamefont{L.}~\bibnamefont{Landau}} \bibnamefont{and}
  \bibinfo{author}{\bibfnamefont{E.}~\bibnamefont{Lifshitz}},
  \bibinfo{journal}{Physik. Z. Sowjetunion} \textbf{\bibinfo{volume}{8}},
  \bibinfo{pages}{153} (\bibinfo{year}{1935}).

\bibitem[{\citenamefont{Abolfath
  et~al.}(2001{\natexlab{b}})\citenamefont{Abolfath, Jungwirth, Brum, and
  MacDonald}}]{abolfath01a}
\bibinfo{author}{\bibfnamefont{M.}~\bibnamefont{Abolfath}},
  \bibinfo{author}{\bibfnamefont{T.}~\bibnamefont{Jungwirth}},
  \bibinfo{author}{\bibfnamefont{J.}~\bibnamefont{Brum}}, \bibnamefont{and}
  \bibinfo{author}{\bibfnamefont{A.~H.} \bibnamefont{MacDonald}},
  \bibinfo{journal}{Phys. Rev. B} \textbf{\bibinfo{volume}{63}},
  \bibinfo{pages}{054418} (\bibinfo{year}{2001}{\natexlab{b}}).

\bibitem[{\citenamefont{Koeder~{\it et al.}}(2003)}]{koeder}
\bibinfo{author}{\bibfnamefont{A.}~\bibnamefont{Koeder~{\it et al.}}},
  \bibinfo{journal}{Appl. Phys. Lett.} \textbf{\bibinfo{volume}{82}},
  \bibinfo{pages}{3278} (\bibinfo{year}{2003}).

\bibitem[{\citenamefont{Sparks}(1970)}]{sparks70c}
\bibinfo{author}{\bibfnamefont{M.}~\bibnamefont{Sparks}},
  \bibinfo{journal}{Phys. Rev B} \textbf{\bibinfo{volume}{1}},
  \bibinfo{pages}{3869} (\bibinfo{year}{1970}).

\bibitem[{\citenamefont{Hoekstra et~al.}(1977)\citenamefont{Hoekstra, van
  Stapele, and Robertson}}]{hoekstra77}
\bibinfo{author}{\bibfnamefont{B.}~\bibnamefont{Hoekstra}},
  \bibinfo{author}{\bibfnamefont{R.~P.} \bibnamefont{van Stapele}},
  \bibnamefont{and} \bibinfo{author}{\bibfnamefont{J.~M.}
  \bibnamefont{Robertson}}, \bibinfo{journal}{J. Appl. Phys.}
  \textbf{\bibinfo{volume}{48}}, \bibinfo{pages}{382} (\bibinfo{year}{1977}).

\bibitem[{\citenamefont{Sparks}(1969)}]{sparks69}
\bibinfo{author}{\bibfnamefont{M.}~\bibnamefont{Sparks}},
  \bibinfo{journal}{Phys. Rev. Lett} \textbf{\bibinfo{volume}{22}},
  \bibinfo{pages}{1111} (\bibinfo{year}{1969}).

\bibitem[{\citenamefont{Soohoo}(1963)}]{soohoo63}
\bibinfo{author}{\bibfnamefont{R.~F.} \bibnamefont{Soohoo}},
  \bibinfo{journal}{Phys. Rev} \textbf{\bibinfo{volume}{131}},
  \bibinfo{pages}{594} (\bibinfo{year}{1963}).

\bibitem[{\citenamefont{Ament and Rado}(1955)}]{Ament55}
\bibinfo{author}{\bibfnamefont{W.~S.} \bibnamefont{Ament}} \bibnamefont{and}
  \bibinfo{author}{\bibfnamefont{G.~T.} \bibnamefont{Rado}},
  \bibinfo{journal}{Phys. Rev. B} \textbf{\bibinfo{volume}{97}},
  \bibinfo{pages}{1558} (\bibinfo{year}{1955}).

\bibitem[{\citenamefont{Pincus}(1960)}]{pincus60}
\bibinfo{author}{\bibfnamefont{P.}~\bibnamefont{Pincus}},
  \bibinfo{journal}{Phys. Rev. B} \textbf{\bibinfo{volume}{118}},
  \bibinfo{pages}{658} (\bibinfo{year}{1960}).

\bibitem[{\citenamefont{Kittel}(1949)}]{kittel49}
\bibinfo{author}{\bibfnamefont{C.}~\bibnamefont{Kittel}},
  \bibinfo{journal}{Revs. Modern Phys.} \textbf{\bibinfo{volume}{21}},
  \bibinfo{pages}{541} (\bibinfo{year}{1949}).

\bibitem[{\citenamefont{Ujsaghy et~al.}(1996)\citenamefont{Ujsaghy, Zawadowski,
  and Gyorffy}}]{ujsaghy}
\bibinfo{author}{\bibfnamefont{O.}~\bibnamefont{Ujsaghy}},
  \bibinfo{author}{\bibfnamefont{A.}~\bibnamefont{Zawadowski}},
  \bibnamefont{and} \bibinfo{author}{\bibfnamefont{B.~L.}
  \bibnamefont{Gyorffy}}, \bibinfo{journal}{Phys. Rev. Lett.}
  \textbf{\bibinfo{volume}{76}}, \bibinfo{pages}{2378} (\bibinfo{year}{1996}).

\bibitem[{\citenamefont{Yu et~al.}(2002{\natexlab{b}})\citenamefont{Yu,
  Walukiewicz, Wojtowicz, Kuryliszyn, Liu, Sasaki, and J.K.}}]{Rutherford}
\bibinfo{author}{\bibfnamefont{K.}~\bibnamefont{Yu}},
  \bibinfo{author}{\bibfnamefont{W.}~\bibnamefont{Walukiewicz}},
  \bibinfo{author}{\bibfnamefont{T.}~\bibnamefont{Wojtowicz}},
  \bibinfo{author}{\bibfnamefont{I.}~\bibnamefont{Kuryliszyn}},
  \bibinfo{author}{\bibfnamefont{X.}~\bibnamefont{Liu}},
  \bibinfo{author}{\bibfnamefont{Y.}~\bibnamefont{Sasaki}}, \bibnamefont{and}
  \bibinfo{author}{\bibfnamefont{F.}~\bibnamefont{J.K.}},
  \bibinfo{journal}{Phys. Rev. B} \textbf{\bibinfo{volume}{65}},
  \bibinfo{pages}{201303} (\bibinfo{year}{2002}{\natexlab{b}}).

\bibitem[{\citenamefont{Gallagher~{\it et al.}}()}]{gallagher}
\bibinfo{author}{\bibfnamefont{B.~L.} \bibnamefont{Gallagher~{\it et al.}}},
  \bibinfo{note}{cond-mat/0307140}.

\bibitem[{\citenamefont{Stone et~al.}()\citenamefont{Stone, Ku, Potashnik,
  Sheu, Samarth, and Schiffer}}]{shiffer-cap}
\bibinfo{author}{\bibfnamefont{M.}~\bibnamefont{Stone}},
  \bibinfo{author}{\bibfnamefont{K.}~\bibnamefont{Ku}},
  \bibinfo{author}{\bibfnamefont{S.~J.} \bibnamefont{Potashnik}},
  \bibinfo{author}{\bibfnamefont{B.~L.} \bibnamefont{Sheu}},
  \bibinfo{author}{\bibfnamefont{N.}~\bibnamefont{Samarth}}, \bibnamefont{and}
  \bibinfo{author}{\bibfnamefont{P.}~\bibnamefont{Schiffer}},
  \bibinfo{note}{cond-mat/0307255}.

\end{thebibliography}
\end{document}